\documentclass{article}
\usepackage{amssymb}
\usepackage{amsmath}
\usepackage[margin=1in]{geometry}
\usepackage{graphicx}

\begin{document}

\title{Light induced stress and work in photomechanical materials}
\author{Anastasiia Svanidze$^{1}$\thanks{%
Email: asvanidz@kent.edu}, Tianyi Guo$^{1}$ \thanks{
Email: tguo2@kent.edu}, Xiaoyu Zheng$^{2}$\thanks{
Email: xzheng3@kent.edu}, Peter Palffy-Muhoray$^{1,2}$\thanks{
Corresponding author. Email: mpalffy@kent.edu} \\
\emph{$^1$Advanced Materials and Liquid Crystal Institute, Kent State
University, OH 44242 USA}\\
\emph{$^2$Department of Mathematical Sciences, Kent State University, OH 44242 USA%
} }
\date{}
\maketitle

\begin{abstract}
Increasingly important photomechanical materials produce stress and mechanical work when illuminated. We propose experimentally accessible performance metrics for photostress and photowork,
enabling comparison of materials performance. We relate these metrics to material properties,  providing a framework for the design and optimization of photomechanical materials.
\end{abstract}

The availability of high intensity light sources and the ease of optical
power transmission have generated considerable interest in materials capable
of directly exerting forces and performing mechanical work under the
influence of light. Such photomechanical (PM) materials exhibit stresses on
illumination, can undergo shape changes and perform mechanical work. Despite
this interest, there are no standard methods and criteria in use today for
assessing materials performance in producing photostress or performing
photowork. Materials of particular interest are liquid crystal elastomers 
\cite{1,2} and networks \cite{3,4}, semicrystalline polymers \cite{5,6} and
organic crystals \cite{7, 8, 9}.

We propose here performance criteria that can be evaluated using results
from experiments and which enable comparison of materials performance. We
further establish the dependence of these criteria on material parameters,
opening the door towards systematic optimization of PM materials.

The first challenge in establishing the relation between the light induced
photostress $\boldsymbol{\sigma }^{ph}$ and the incident light is the
identification of an appropriate descriptor of the light. PM materials are
typically anisotropic, with a polarization, intensity and wavelength
dependent photoresponse. In addition to intensity, the descriptor thus needs
to carry information about polarization and frequency; it also needs to be a
tensor.

Although interestingly the Minkowski stress does not carry polarization
information, its first term, the outer product of the electric field and the
electric displacement of light, does, and hence can serve in this role. We
denote it as $\boldsymbol{\sigma }^{l}=\mathbf{ED}$, the light stress. This
is simply related to the light intensity $I=\frac{1}{2}c\text{Tr}(%
\boldsymbol{\sigma }^{l})$, where $c$ is the speed of light, and Tr is the
trace.

In the linear regime, one expects the constitutive relation 
\begin{equation}
\boldsymbol{\sigma }^{ph}=\mathbf{R}\boldsymbol{\sigma }^{l},  \label{def}
\end{equation}%
where $\mathbf{R}$ is a $4^{th}$ rank tensor representing response, the
stress (or momentum current density) gain. The dimensionless elements of $%
\mathbf{R}$ quantify the efficiency of PM materials in directly producing
stress from light, and enable comparison of materials performance. For
materials with high symmetry, such as transversely isotropic materials, $%
\mathbf{R}$ contains relatively few independent elements whose experimental
determination is feasible.

If the stresses are symmetric and the sample is uniaxial, one can write%
\begin{equation}
\left[ 
\begin{array}{c}
\sigma _{11}^{ph} \\ 
\sigma _{22}^{ph} \\ 
\sigma _{33}^{ph} \\ 
\sigma _{23}^{ph} \\ 
\sigma _{13}^{ph} \\ 
\sigma _{12}^{ph}%
\end{array}%
\right] =\left[ 
\begin{array}{cccccc}
R_{11} & R_{12} & R_{13} & 0 & 0 & 0 \\ 
R_{21} & R_{22} & R_{23} & 0 & 0 & 0 \\ 
R_{31} & R_{32} & R_{33} & 0 & 0 & 0 \\ 
0 & 0 & 0 & R_{44} & 0 & 0 \\ 
0 & 0 & 0 & 0 & R_{55} & 0 \\ 
0 & 0 & 0 & 0 & 0 & R_{66}%
\end{array}%
\right] \left[ 
\begin{array}{c}
\sigma _{11}^{l} \\ 
\sigma _{22}^{l} \\ 
\sigma _{33}^{l} \\ 
\sigma _{23}^{l} \\ 
\sigma _{13}^{l} \\ 
\sigma _{12}^{l}%
\end{array}%
\right],
\end{equation}%
and by using light linearly polarized along a principal axis, the diagonal
elements of $\mathbf{R}$ can be determined.

We next inquire about the connection between the elements of $\mathbf{R}$
and material parameters of PM samples.

\begin{figure}[htb]
\centering
\includegraphics[height=0.3\textwidth]{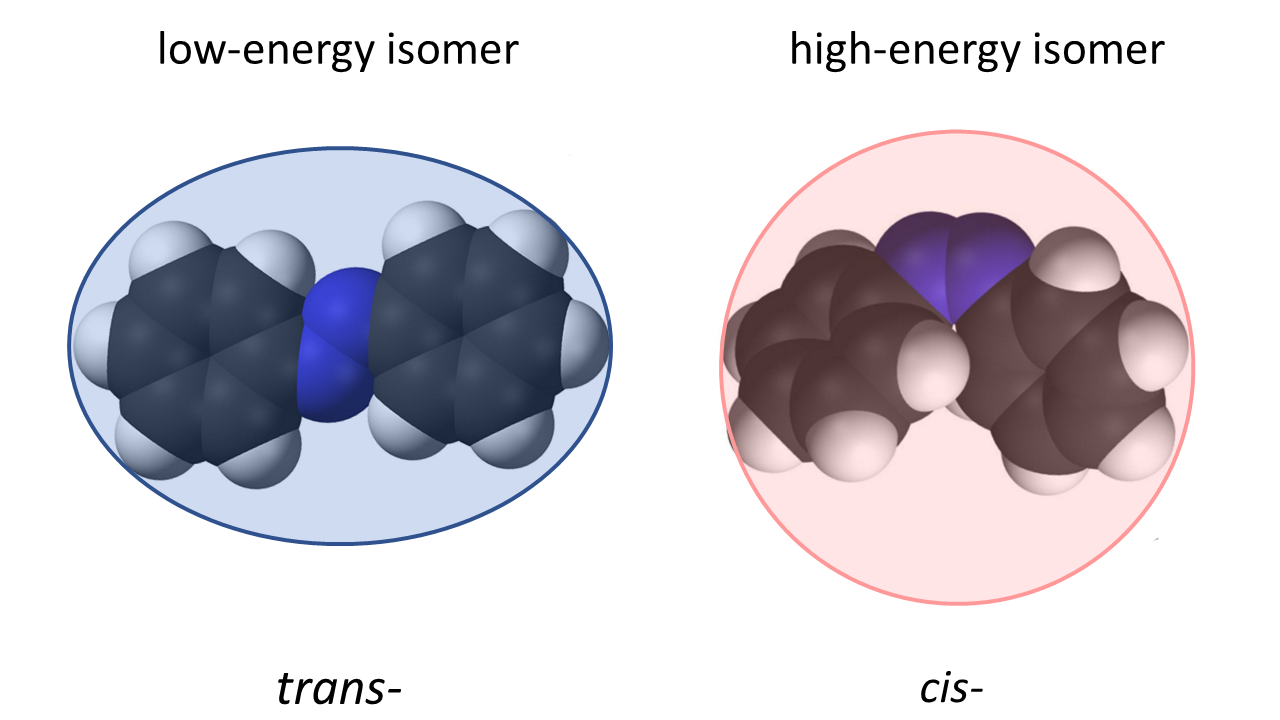}
\caption{A schematic shows the shape change of a dye isomer under
illumination and is excited from the low energy ``trans" (left) to the high
energy ``cis" (right) conformation. The bare strain tensor can be calculated
based on the change of semiaxes of the ellipsoid.}
\label{fig:Fig2}
\end{figure}

To gain some insight, we make use of the following elementary model. We
assume that PM bulk materials consist of a solid elastic host, in which
discrete photoisomerizable `dye' molecules with volume $v_{d}$ are embedded
randomly but with uniform average density. When illuminated, a dye molecule
in the low-energy isomer conformation absorbs a single photon with energy $%
h\nu _{d}$, and undergoes a prescribed volume-conserving shape-changing
transformation, such as from a low-energy prolate ellipsoid to a sphere, the
high-energy isomer conformation. The dye shape change can be described by a
bare dye strain $\boldsymbol{\varepsilon }_{0}^{d}$, see Fig.\ \ref{fig:Fig2}%
. In the absence of an externally applied stress, in an ideal continuum
model, where identical solid inclusions undergo a volume-conserving strain $%
\boldsymbol{\varepsilon }_{0}^{d}$ in an elastic solid host, the resulting
bulk strain may be assumed to be $\boldsymbol{\varepsilon }^{ph}=\phi _{hd}%
\boldsymbol{\varepsilon }_{0}^{d}$, where $\phi _{hd}$ is the volume
fraction of the high-energy conformers. However, the orientations of the dye
molecules are not identical; the relevant quantity is their orientational
average $\langle\boldsymbol{\varepsilon }_{0}^{d}\rangle$ $=S_{d}\boldsymbol{\varepsilon 
}_{0}^{d}$ where $-\frac{1}{2}\leq S_{d}\leq 1$ is the orientational order
parameter. In addition, interactions with the host can further modify this%
\footnote{%
In the simplest case, $\alpha =$ $1$ but in general, it can be reduced or
increased due to the interaction of the dye and the host. Examples would be
free volume effects in the host reducing the effective bare dye strain, or
order parameter change of a liquid crystal host, reducing anisotropic
dispersion interaction due to photoisomerization of the dye; coupling of the
order parameter to the polymer network results in bulk strain. Ref. \cite%
{Dunn} reports vales of $\varepsilon ^{d}>1$.}, and we write instead $%
\boldsymbol{\varepsilon }^{ph}=\phi _{hd}\boldsymbol{\varepsilon }^{d}$,
where $\boldsymbol{\varepsilon }^{d}=\alpha S_{d}\boldsymbol{\varepsilon }%
_{0}^{d}$ is the effective dye strain. The corresponding photostress $%
\boldsymbol{\sigma }^{ph}$, measured at zero bulk strain, is assumed to be 
\begin{equation}
\boldsymbol{\sigma }^{ph}=\phi _{hd}\mathbf{C}:\boldsymbol{\varepsilon }^{d},
\label{sigma_ph}
\end{equation}%
where $\mathbf{C}$ is the stiffness tensor of the host.

The photoinduced high-energy conformers relax back into the low-energy
conformation after a lifetime $\tau $. For simplicity, we assume that the
high-energy conformer always has enough energy to guarantee the strain $%
\boldsymbol{\varepsilon }^{d}$, and we assume that $\tau $ is a constant
regardless of the environment. In the photostationary state, the decay rate
of the high energy isomer population must equal the creation rate from the
low energy isomer population with volume fraction $\phi _{ld}$, and 
\begin{equation}
\frac{\phi _{hd}}{\phi _{ld}}=(\boldsymbol{\sigma }^{d})^{-1}:\boldsymbol{%
\sigma }^{l},  \label{c}
\end{equation}%
where 
\begin{equation}
(\boldsymbol{\sigma }^{d})^{-1}=\frac{c\tau }{h\nu _{d}}\mathbf{a}^{ld},
\end{equation}%
and $\mathbf{a}^{ld}$ is the absorption cross-section tensor of the
low-energy conformer.\footnote{%
We take the this to be%
\begin{equation}
\mathbf{a}^{ld}=\frac{2\pi }{\varepsilon _{0}\lambda }\text{Im}\mathbf{%
\alpha }^{d},
\end{equation}%
where $\text{Im}\mathbf{\alpha }^{ld}$ is the imaginary part of the
polarizability, $\varepsilon _{0}$ is the permittivity of free space, and $%
\lambda $ is the wavelength.} We note that dye stress $\boldsymbol{\sigma }%
^{d}$, with the units of energy current density, is a material property.
Since the total dye volume fraction $\phi _{d}=\phi _{hd}+\phi _{ld}$ is a
constant, 
\begin{equation}
\phi _{hd}=\phi _{d}\frac{(\boldsymbol{\sigma }^{d})^{-1}:\boldsymbol{\sigma 
}^{l}}{1+(\boldsymbol{\sigma }^{d})^{-1}:\boldsymbol{\sigma }^{l}},
\label{nl}
\end{equation}%
indicating nonlinearity and saturation when $(\boldsymbol{\sigma }^{d})^{-1}:%
\boldsymbol{\sigma }^{l}\gtrsim 1$.

The photostress, in the linear regime, from Eqs.\ \eqref{sigma_ph} and %
\eqref{c}, is 
\begin{equation}
\boldsymbol{\sigma }^{ph}=\phi _{ld}\mathbf{C}:\boldsymbol{\varepsilon }%
^{d}((\boldsymbol{\sigma }^{d})^{-1}:\boldsymbol{\sigma }^{l}),
\label{sigma_ph_2}
\end{equation}%
and the stress gain function $\mathbf{R}$, from Eqs.\ \eqref{def} and %
\eqref{sigma_ph_2}, is 
\begin{equation}
\mathbf{R}=\phi _{ld}\mathbf{C}:\boldsymbol{\varepsilon }^{d}(\boldsymbol{%
\sigma }^{d})^{-1}.  \label{R}
\end{equation}

The dependence of the stress gain $\mathbf{R}$ on material parameters is a
key result. Its constituents are more readily seen in an isotropic system;
there Eq.\ \eqref{R} can be written as $R=\phi _{ld}\varepsilon ^{d}Yc/I_{0}$%
, where $Y$ is Young's modulus, and $I_{0}=h\nu _{d}/a^{ld}\tau .$ According
to Eq.\ \eqref{R}, our model indicates that, for a given light intensity,
photostress is proportional to the corresponding elastic modulus \cite{five}
of PM materials.

The above result can be compared with experiment. By illuminating samples
with light polarized along the symmetry axis $\mathbf{\hat{z}}$ of uniaxial
liquid crystal elastomers and other transversely isotropic samples, the
element $R_{zz}=\sigma _{zz}^{ph}c/I$ can be determined. Measuring stress
gain in bels, the quantity $R_{zz}^{\prime }=\log _{10}R_{zz}$, as a
function of Young's modulus, is shown in Fig.\ \ref{fig:Fig1} below,
demonstrating the experimentally observed stiffness dependence of the
response. 

\begin{figure}[htb]
\centering
\includegraphics[height=0.5\textwidth]{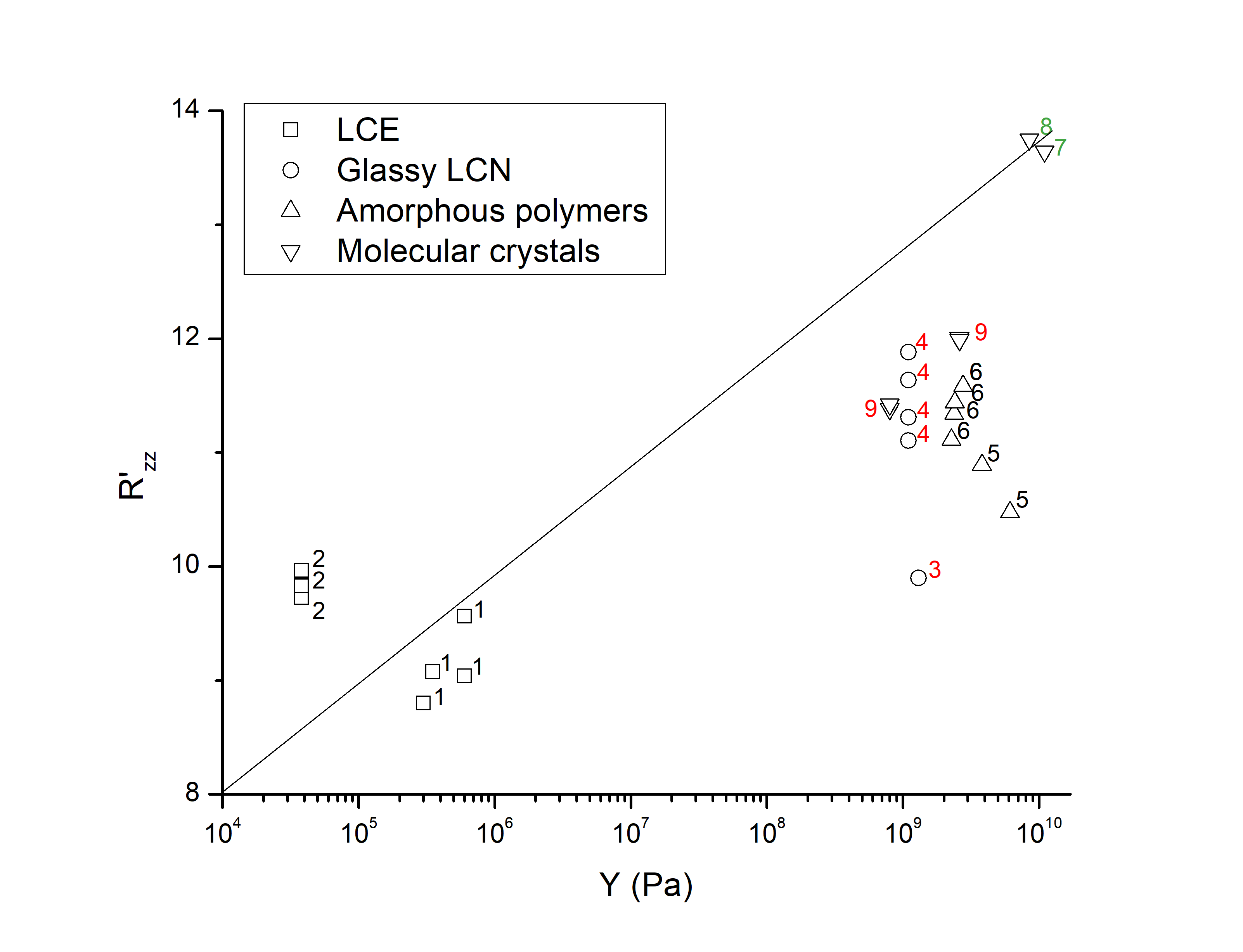}
\caption{Graph of $R_{zz}^{\prime }$ vs.\ Young's modulus $Y$. The numbers
on the graph indicates that $R_{zz}^{\prime }$ is calculated based on data
from the corresponding article in the reference list. The line indicates proportionality;
deviations are ascribed to differences in values of $\phi _{ld},\varepsilon
^{d}$ and $I_{0}$.}
\label{fig:Fig1}
\end{figure}

We next consider the efficiency of PM materials in converting light energy
to mechanical work. We define the efficiency $\Gamma $ as the ratio of
mechanical work done on a load to the energy absorbed from light.
Experimental determination of $\Gamma $ is usually straightforward.  To
relate the efficiency to material parameters, we make use of the relation
between photostress and light in Eqs.\ \eqref{def} and \eqref{R}.

When a PM material is illuminated, photoisomerizable dye molecules
photoisomerized into the high-energy conformation exhibit a prescribed
strain $\boldsymbol{\varepsilon }^{d}$, leading to bulk photostress and
photostrain. If there are no externally applied stresses, the sample
deforms, and mechanical work produced by light is stored as elastic energy.
If an external load is present providing an external stress, then the
mechanical work enabled by light is partitioned between work on the load and
elastic energy stored in the PM host. To simply illustrate this, we consider
an illuminated isotropic elastomer host without any external stress, but
with a volume conserving uniaxial photostrain and corresponding
uniaxial photostress. The stored elastic energy density in a volume
conserving material is then 
\begin{equation}
\mathcal{E}_{st}=\frac{3}{2}G(\varepsilon^{ph} _{zz})^{2}=\frac{3}{8}\frac{1}{G}%
(\sigma _{zz}^{ph})^{2},
\end{equation}%
where $G$ is the shear modulus and $\varepsilon _{zz}$ is element of the
bulk strain. If a load is present, with a single stress component $\sigma
_{zz}^{load}$, the stored energy becomes 
\begin{equation}
\mathcal{E}_{st}=\frac{3}{8}\frac{1}{G}(\sigma _{zz}^{ph}-\frac{2}{3}\sigma
_{zz}^{load})^{2}.
\end{equation}%
If the load stress is written in terms of the photostress as $\sigma
_{zz}^{load}=\gamma \sigma _{zz}^{ph}$, where $\gamma <1$, the stored
elastic energy density becomes 
\begin{equation}
\mathcal{E}_{st}=\frac{1}{2}Y(\varepsilon _{zz}^{ph})^{2}(1-\frac{2}{3}%
\gamma )^{2},
\end{equation}%
where $Y=3G$ is Young's modulus, while the work density $\mathcal{W}$ done
on the load is%
\begin{equation}
\mathcal{W}=\sigma _{zz}^{load}\varepsilon _{zz}=\frac{1}{2}Y(\varepsilon
_{zz}^{ph})^{2}\frac{4}{3}\gamma (1-\frac{2}{3}\gamma ).
\end{equation}%
The photowork done on the external load thus depends on the ratio $\gamma $
of the load stress to the photostress; it is maximized here when an
impedance match function, here $f(\gamma )=\frac{4}{3}\gamma (1-\frac{2}{3}%
\gamma )$, is a maximum. When $\gamma=3/4$, $\sigma^{load}_{zz}=3/4\sigma^{ph}_{zz}=1/2Y\varepsilon^{ph}_{zz}$.

In general, the mechanical work density done by the light on the sample is 
\begin{equation}
\mathcal{W}=\frac{1}{2}\phi _{hd}^{2}\boldsymbol{\varepsilon }^{d}:\mathbf{C}%
:\boldsymbol{\varepsilon }^{d}f(\gamma ),  \label{a}
\end{equation}%
where $f(\gamma )$ may differ from the expression above, but we expect its
argument $\gamma $ to remain the same (the ratio of the load stress to the
photostress), and its maximum to occur when $\gamma $ is of order unity.

The system is assumed to be overdamped; the time for work is taken to be the
characteristic response time $t_{r}$ of the system.\footnote{%
For simplicity, we don't pursue the anisotropic viscoelastic origins of $%
t_{r}$ here.} The energy input per volume from the absorbed light in time $%
t_{r}$ is 
\begin{equation}
\mathcal{E}_{in}=\frac{\phi _{ld}}{v_{d}}c\mathbf{a}^{ld}:\boldsymbol{\sigma 
}^{l}t_{r},  \label{b}
\end{equation}%
where $v_{d}$ is the volume of a dye molecule.

The efficiency, $\Gamma =$ $\mathcal{W}/\mathcal{E}_{in}$, from Eqs.\ %
\eqref{c}, \eqref{a} and \eqref{b}, is 
\begin{equation}
\Gamma =\frac{\frac{1}{2}\boldsymbol{\varepsilon }^{d}:\phi _{hd}\mathbf{C}:%
\boldsymbol{\varepsilon }^{d}}{(t_{r}/\tau )(h\nu _{d}/v_{d})}f(\gamma ).
\label{eff2}
\end{equation}

The dependence of the light-to-work efficiency $\Gamma $ on material
parameters is our second key result. In this form, Eq.\ \eqref{eff2}
enables straightforward interpretation. The quantity $\frac{1}{2}\boldsymbol{%
\varepsilon }^{d}:\phi _{hd}\mathbf{C}:\boldsymbol{\varepsilon }^{d}$  in
Eq.\ \eqref{eff2} is the work density of a single dye molecule in undergoing
strain $\boldsymbol{\varepsilon }^{d}$ in a medium with effective stiffness
constant $\phi _{hd}\mathbf{C}$ due to photoisomerization. The quantity ($%
h\nu _{d}/v_{d})\,$in the\ denominator is the energy density of the
photoisomerized dye molecule; $(t_{r}/\tau )$ is the number of photons
absorbed in doing the work.  The last term $f(\gamma )$, related to
impedance match, is the fraction of mechanical energy that has gone into
work on the load. The volume fraction $\phi _{hd}$ depends on illumination
as given in Eq.\ \eqref{nl}, however, it cannot be greater than $\phi _{d}$.

Lack of reliable parameters values makes comparison of experiment and model
predictions challenging. Data on high-energy isomer lifetimes $\tau $ and
low-energy isomer absorption cross-sections is scarce. Information regarding
the effective dye strains $\boldsymbol{\varepsilon }^{d}$ is not readily
available. Although the bare dye strain $\mathbf{\varepsilon }_{0}^{d}$ can
be reasonable estimated, from molecular models, the effective dye strain $%
\mathbf{\varepsilon }^{d}=\alpha S_{d}\mathbf{\varepsilon }_{0}^{d}$ is
more difficult to estimate since there is little data available on the
orientational order parameter $S_{d}$ and even less on the coupling term $%
\alpha $. Nonetheless, Eq.\ \eqref{eff2}, together with Eq.\ \eqref{nl}, gives an
explicit expression for efficiency in terms of measurable material
parameters and illumination. 

To enable rough comparison with experiments, we assume that one-half of the
dye molecules are in the high-energy conformation, $\phi _{hd}=\phi _{d}/2$,
and $\varepsilon ^{d}=2\varepsilon ^{ph}/\phi _{d}$, and  calculating the
molecular volume of dyes from mass density and assuming that $f=1/2$ and $t_{r}=\tau$, we estimate the efficiency using%
\begin{equation}
\Gamma _{th}\simeq \frac{\frac{1}{2}(\varepsilon ^{ph})^{2}Y}{\phi _{d}(h\nu
_{d}/v_{d})}.  \label{fom}
\end{equation}%
Using data from publications with adequate photowork information, we provide
experimental and theoretical efficiencies for two different liquid crystal
elastomers with azo dye and for a two-component organic diarylethene crystal
in Table \ref{tbl1}. 
\begin{table}[tbp]
\caption{Photowork efficiencies of three photomechanical materials.}
\label{tbl1}\centering
\begin{tabular}{llllllll}
\hline\hline
Ref. & Material  & $\varepsilon ^{ph}$ & $\phi _{d}$ & $Y(Pa)$ & $h\nu _{d}/v_{d}(Pa)$
& $\Gamma _{ex}$ & $\Gamma _{th}$ \\ \hline
\cite{PPM} & LCE & $6\times 10^{-2}$ & ${\normalsize 1\times 10}^{-3}$ & $%
{\normalsize 1\times 10}^{4}$ & $4\times 10^{8}$ & $4\times 10^{-6}$ & $%
4\times 10^{-5}$ \\ \hline
\cite{Ikeda Yu} & LCN & $8\times 10^{-3}$ & $1$ & ${\normalsize 3\times 10}
^{8}$ & $6\times 10^{8}$ & ${\normalsize 1\times 10}^{-5}$ & $6\times
10^{-6} $ \\ \hline
\cite{Morimoto} & DAE Crystal & $2\times 10^{-3}$ & $5\times 10^{-1}$ & $%
1\times 10^{10}$ & $5\times 10^{8}$ & $7\times 10^{-5}$ & $8\times 10^{-5}$
\\ \hline\hline
\end{tabular}%
\end{table}

This suggests reasonable agreement between experiment and model predictions,
given range of material properties and the assumptions made in lieu of
missing data. Relevant data to enable more detailed comparison of experiment
efficiency and model predictions can be obtained from experiments.\footnote{%
The low-energy isomer absorption cross-section $\mathbf{a}$ can be obtained
from the absorption coefficient/decay length of the dyed sample. The
high-energy isomer lifetime $\tau $ can be obtained from the time dependence
of the absorption spectrum of the sample. This allows determination of the
characteristic intensity; then the from the total dye volume fraction $\phi
_{d}$, the volume fraction $\phi _{hd}$ of the high-energy isomer can be
obtained as function of incident intensity. Once the volume fraction $\phi
_{hd}$ is known, the dye strain can be calculated from the measured
photoinduced bulk strain. Knowing the photostress from experiment, the
load-to-photostress ratio $\gamma $ can be calculated.}

Eq.\ \eqref{eff2} clearly identifies the factors contributing to the
efficiency. The dominant contribution to work is $\alpha S_{d}\mathbf{%
\varepsilon }_{0}^{d}:\mathbf{C}:\alpha S_{d}\mathbf{\varepsilon }_{0}^{d}$,
 where the joint contribution of all terms needs to be maximized. 
Although this is challenging, the contributing mechanisms have been
identified. The volume fraction of the high energy isomer given by Eq.\ \eqref%
{nl} is critical, since due to absorption $\boldsymbol{\sigma }^{l}$ in
thick samples can be small, resulting in dramatic reduction of $\phi _{hd}$
\cite{Warner1, Warner2}. The energy density of the high energy dye
is given; it represents the maximum energy available for work. Matching
system response time to the high energy conformer lifetime is desirable, as
is matching load stress to photostress.  Current experimental results
indicate that bulk of the absorbed light energy goes into heating the
sample.  Separating thermal and photoinduced contributions to mechanical
work are therefore of paramount importance when probing the performance of
PM materials.

In the light of the above, we propose, as a basic figure of merit, the ratio
of work and energy densities from Eq.\ \eqref{fom}:%
\begin{equation}
FOM=\frac{\frac{1}{2}(\varepsilon ^{ph})^{2}Y}{\phi _{d}(h\nu _{d}/v_{d})},
\end{equation}%
which can be readily determined.  We mention that results from a different
but compatible model are consistent with ours presented here \cite{Kuzyk}; a
detailed comparison will be presented elsewhere. Details of the
photomechanical coupling in nematic elastomers have also been studied,
including the question of mechanical stability \cite{Bai}.

In summary, we have identified the appropriate descriptor of the
illuminating light, the light stress $\mathbf{\sigma }^{l}=\mathbf{ED}$. 
We have provided a simple model, on the basis of which we have proposed two
criteria for assessing the performance of a photomechanical material: the
stress gain $\mathbf{R}$, and the work efficiency $\mathbf{\Gamma }$.  Both
can be readily determined from experimental data.  The stress gain is a
novel criterion which measures the ratio of light induced stress in the bulk
material relative to the stress from light, and the work efficiency $\Gamma $
measures the ratio of work done on an external load relative to the energy
absorbed per dye. We have related both of these quantities to material
properties, and showed reasonable agreement between model predictions and
experiment.  We have identified the material properties of dye stress $%
\mathbf{\sigma }^{d}$ and absorption cross-section tensor $\mathbf{a}^{ld}$
and indicated their role, and indicated how they may be determined from
measurements.  

Most importantly, based on insights gained from the model, we identified the
individual processes and their contributions to the efficiency of light
doing mechanical work. Accordingly, we have indicated strategies to optimize
performance related to these processes, and provide direction for improving
PM materials performance.

Acknowledgements:  We acknowledge valuable discussions with D.\ Broer, Eindhoven University of Technology) and T.\ Ikeda (Chuo University). We are grateful for collaboration on
materials preparation with S-W.\ Oh and for useful discussions with R.\
Hayward (Colorado) and the MURI team. We are grateful for literature
references provided by W.\ Xu. This work was supported by the Office of Naval
Research [ONR N00014-18-1-2624].


\begin{thebibliography}{99}
\bibitem{1} A. S\'{a}nchez-Ferrer, A. Merekalov and H. Finkelmann.
Opto-mechanical effect in photoactive nematic side-chain liquid-crystalline
elastomers. Macromolecular rapid communications, 32(8), 671-678 (2011).

\bibitem{2} C. L. M. Harvey and E. M. Terentjev E. M. Role of polarization
and alignment in photoactuation of nematic elastomers. The European Physical
Journal E, 23(2), 185-189 (2007).

\bibitem{3} C. L. van Oosten, K. D. Harris, C. W. M \ Bastiaansen and D.J.
Broer. Glassy photomechanical liquid-crystal network actuators for
microscale devices. The European Physical Journal E, 23(3), 329-336 (2007).

\bibitem{4} C. L. van Oosten, D. Corbett, D. Davies, M. Warner,C. W.
Bastiaansen and D.J. Broer. Bending dynamics and directionality reversal in
liquid crystal network photoactuators. Macromolecules, 41(22), 8592-8596
(2008).

\bibitem{5} K. M. Lee, D. H. Wang, H. Koerner, R. A. Vaia, L. S. Tan, and T.
J. White. Enhancement of photogenerated mechanical force in
azobenzene-functionalized polyimides. Angewandte Chemie International
Edition, 51(17), 4117-4121 \ (2012).

\bibitem{6} J.J. Wie, D.H. Wang, K.M. Lee, L.S. Tan,\& T.J. White. Molecular
engineering of azobenzene-functionalized polyimides to enhance both
photomechanical work and motion. Chemistry of Materials, 26(18), 5223-5230
(2014).

\bibitem{7} M. Morimoto and M. Irie. A diarylethene cocrystal that converts
light into mechanical work. Journal of the American Chemical Society,
132(40), 14172-14178 (2010).

\bibitem{8} F. Terao, M. Morimoto and M. Irie. Light-driven
molecular-crystal actuators: rapid and reversible bending of rodlike mixed
crystals of diarylethene derivatives. Angewandte Chemie International
Edition, 51(4), 901-904 (2012).

\bibitem{9} H. Koshima, R. Matsuo, M. Matsudomi, Y. Uemura ande M. Shiro.
Light-driven bending crystals of salicylidenephenylethylamines in
enantiomeric and racemate forms. Crystal growth \& design, 13(10), 4330-4337
(2013).

\bibitem{five} S-W Oh, T. Guo, A. Kuenstler, R. Hayward, P. Palffy-Muhoray,
X. Zheng, Measuring the five elastic constants of a nematic liquid crystal
elastomer, Liq. Cryst., DOI: 10.1080/02678292.2020.1790680 (2020).

\bibitem{Dunn} M. Dunn, Photomechanics of mono- and polydomain liquid
crystal elastomer films, J. App. Phys. 102, 013506 (2007).

\bibitem{PPM} M. Camacho-Lopez, H. Finkelmann, P. Palffy-Muhoray, and M.
Shelley, Fast liquid-crystal elastomer swims into the dark, Nat Mater 3, 307
(2004).

\bibitem{Ikeda Yu} M. Yamada, M. Kondo, J. I. Mamiya, Y. L. Yu, M.
Kinoshita, C. J. Barrett, and T.Ikeda, Photomobile polymer materials:
Towards light-driven plastic motors, Angew Chem Int Edit 47, 4986 (2008).

\bibitem{Morimoto} M. Morimoto and M. Irie, A Diarylethene Cocrystal that
Converts Light into Mechanical Work, J Am Chem Soc 132, 14172 (2010).

\bibitem{Kuzyk} B. Zhou, E. Bernhardt, A. Bhuyan, Z. Ghorbanishiadeh, N.
Rasmussen, J. Lanska, and M. G. Kuzyk, Theoretical and experimental studies
of photomechanical materials, J. Opt. Soc Am B, 36, 1492-1517 (2019).

\bibitem{Bai} R. Bai and K. Bhattacharya, Photomechanical Coupling in
Photoactive Nematic Elastomers, J. Mech. Phys. Solids 144,104115 (2020).

\bibitem{Warner1} D. Corbett and M. Warner, Linear and Nonlinear
Photoinduced Deformations of Cantilevers, Phys. Rev. Lett. 99, 174301 (2007)

\bibitem{Warner2} D. Corbett,C. Xuan and M. Warner, Deep optical penetration
dynamics in photobending, Phys. Rev. E 92, 013205 (2015)
\end{thebibliography}
\end{document}